\documentclass[pra]{revtex4}
\usepackage{graphicx}
\usepackage{indentfirst}
\usepackage{dcolumn}
\usepackage{amsfonts}
\usepackage{amsmath}
\usepackage{bm}
\newcommand{\ket}[1]{\left|#1\right\rangle}
\newcommand{\bra}[1]{\left\langle#1\right|}
\newcommand{\bgeq}{\begin{equation}}
\newcommand{\edeq}{\end{equation}}
\newcommand{\bgeqn}{\begin{eqnarray}}
\newcommand{\edeqn}{\end{eqnarray}}

\newtheorem{theorem}{Theorem}
\newtheorem{proposition}[theorem]{Proposition}

\begin{document}

\title{Entropic uncertainty relation for mutually unbiased bases}

\author{Shengjun Wu$^{1,2}$}
\author{Sixia Yu$^{2,3}$}
\author{Klaus M{\o}lmer$^1$}%
\affiliation{
$^1$The Lundbeck Foundation Theoretical Center for Quantum System Research \\
Department of Physics and Astronomy, Aarhus University, Denmark
\\
$^2$Hefei National Laboratory for Physical Sciences at Microscale and Department of Modern Physics \\
 University of Science and Technology of China, Hefei, Anhui 230026, P. R.
 China \\
$^3$Centre for Quantum Technologies and Physics Department \\
National University of Singapore, 2 Science Drive 3, Singapore 117542
}

\date{\today}

\begin{abstract}
We derive new inequalities for the probabilities of projective
measurements in mutually unbiased bases of a qudit system. These
inequalities lead to wider ranges of validity and tighter bounds on
entropic uncertainty inequalities previously derived in the
literature.
\end{abstract}

\maketitle

\section{Introduction}

Heisenberg's position-momentum uncertainty relation led to Bohr's
introduction of the complementarity principle, which limits the joint measurability,
or knowability, of different properties of a physical system.
Complementarity is profoundly linked with the Copenhagen interpretation of quantum theory,
according to which it poses limitations on a physical system's ability to
manifest certain physical properties and, hence, on the meaning of physical reality of
these properties. At a more quantitative level, the nonexistence of a basis for
a Hilbert space whose basis states are simultaneous eigenstates of two non-commuting
observables leads to a formal relationship between statistical predictions possible
for measurement outcomes of such observables on a quantum system. The standard
deviations of any two Hermitian operators $\Omega_1$ and $\Omega_2$ on a finite-dimensional
system Hilbert space, defined as
$\Delta \Omega_i = \sqrt{\langle \Omega_i^2\rangle-\langle \Omega_i\rangle^2}$ obey
the Robertson-Schr\"odinger uncertainty inequality
$\Delta \Omega_1 \Delta \Omega_2 \geq \frac{1}{2}\sqrt{|\langle[\Omega_1,\Omega_2]\rangle|^2+
|\langle \{(\Omega_1-\langle \Omega_1\rangle),(\Omega_2-\langle \Omega_2\rangle)\}
\rangle|^2}$.
The inequality with only the commutator term is due to Robertson \cite{Robertson29},
while the tighter bound with the anti-commutator term included was given by
Schr\"odinger \cite{Schroedinger30}. Quantum theory is applied to provide theoretical
predictions in the form of expectation values, and the quantum mechanical uncertainties
play an important role both in the comparison between theory and experiments and in the
assessment of the possible use of simpler, e.g., semiclassical, theoretical methods.

In quantum information theory, complementarity and quantum mechanical uncertainty
are central concepts because they provide the ultimate limits on how much information
can be extracted by measurements on a physical system. Thus, on the one hand, the
uncertainty relation quantitatively limits the achievements of computing and
communication systems, and on the other hand, it provides security against adversary
attacks on a secret communication system. In quantum information theory it is not
the magnitude of physical observables that is of interest, but to a much larger extent
binary values corresponding to the identification of a state being occupied with zero
or unit occupancy.  When projective measurements are carried out to determine
if a quantum system is in a particular basis state, the resulting average population
is identified as a weighted sum of the measurement outcomes zero and unity, and since
the  projection operators on non-orthogonal states are non-commuting observables,
the population of such states obeys uncertainty relations. It is in this context
particularly relevant to consider the so-called mutually unbiased bases
(MUBs) \cite{Ivanovic81,wootters89,KR03,Pittenger04}, which are defined by the property
that the squared overlaps between a basis state in one basis and all basis states
in the other bases are identical, and hence the detection of a particular basis state
does not give away any information about the state if it was prepared in another basis.
The original quantum cryptography protocol by Bennett and Brassard \cite{BB84},
with photons polarized along different sets of directions, and the later six-state
protocol \cite{Bruss98,BG98} exactly make use of the indistinguishability of states
within MUBs.

In connection with information theory, the uncertainty relations may, as originally proposed by Deutsch \cite{deutsch83}, be reformulated in terms of entropies, and it is the purpose of the present article to derive uncertainty relations obeyed by entropies for mutually unbiased bases.

In Sec. II, we review some recently derived entropic uncertainty relations for
mutually unbiased bases. In Sec. III, we present two new mathematical results for
the probabilities to measure certain basis vector states on a qudit ($d$-level) quantum
system. In Sec. IV, we present a number of new entropic uncertainty relations
following from our mathematical results, and in Sec. V we conclude with a brief outlook.

\section{Mutually unbiased bases and entropic uncertainty relations}

In a Hilbert space of finite dimension $d$, it is possible to identify mutually unbiased bases,
but except for special cases, it is currently not known how many such bases exist.
If $d$ is a power of a prime $d=p^k$,
there exist $d+1$ mutually unbiased bases, as exemplified by the three bases
corresponding to the three orthogonal
coordinate axes in the Bloch sphere representation of the qubit. For higher values of $d$ it is only generally known that
at last three mutually unbiased bases can always be identified, and it is a topic of ongoing research to search for more
bases in, e.g., the lowest dimension, $d=6$, which is not a power of a prime \cite{BBELTZ}.

In this section we will briefly summarize the results known about entropic uncertainty
relations for MUBs.

For two incompatible observables, defined to have eigenstates which constitute a
pair of MUBs, an entropic uncertainty relation was conjectured by Kraus \cite{kraus87}
and was soon thereafter proved by Maassen and Uffink \cite{MU88}.
This relation can be expressed as follows
\bgeq
H\{p_{i_1}; i\} +H\{p_{i_2}; i\} \geq \log_2 d  , \label{uncer1}
\edeq
where $d$ is the dimension of the system and the Shannon entropy
$H\{p_{i_m}; i\} \equiv \sum_{i=1}^{d} - p_{i_m} \log_2 p_{i_m}$,
with $p_{i_m} = \bra{i_m} \rho \ket{i_m}$ being the probability of obtaining the $i$th result when the state
$\rho$ of a $d$-dimensional system is projected onto the $m$th basis ($m=1,2$).
Eq. (\ref{uncer1}) constitutes an {\sl information exclusion principle} with application
in quantum communication,
which may be readily adapted to take into account inexact measurements and added
noise \cite{hall95,hall97}.

If we assume the existence of $M$ MUBs, we can show that
\bgeq
\sum_{m=1}^{M} H\{p_{i_m}; i\} \geq \frac{M}{2} \log_2 d .  \label{uncer2}
\edeq
If $M$ is even, the result follows from (\ref{uncer1}) by grouping the MUBs in pairs.
If $M$ is odd, we can write the contribution from all basis states twice and make a new grouping of all bases and use (\ref{uncer1}) on the resulting pairs of different MUBs.

If the Hilbert space dimension is a square $d=r^2$ , Ballester and Wehner \cite{BW07}
have shown that inequality (\ref{uncer2}) is tight when $M$ does not exceed the
maximal number of MUBs that exist for an $r$-dimensional system. By tight it is meant
that a quantum state exists and is explicitly given by Ballester and Wehner,
in which the equality sign holds in (\ref{uncer2}).

When the dimension of the system $d$ is a power of a prime, $d+1$ MUBs exist and the entropic uncertainty relation for
all MUBs,
\bgeq
\sum_{m=1}^{d+1} H\{p_{i_m};i\}
\geq \left\{
\begin{array}{ll}
(d+1) \log_2 (\frac{d+1}{2}) &  \textrm{when $d$ is odd,} \\
\frac{d}{2} \log_2 (\frac{d}{2}) + (\frac{d}{2}+1) \log_2 (\frac{d}{2}+1) & \textrm{when $d$ is even.}
\end{array}
\right\}, \label{uncer3}
\edeq
was obtained by Ivanovic \cite{Ivanovic92} and Sanchez-Ruiz \cite{sanchez93,sanchez-ruiz95}.

We shall now proceed to confirm, generalize, and extend the domain of validity of some of the results summarized above.

\section{Two new inequalities}

The derivation of the best entropic uncertainty relation (\ref{uncer3}) for $d+1$ MUBs
when $d$ is a power of a prime is based on the equality
$\sum_{m=1}^{d+1} \sum_{i=1}^d p_{i_m}^2 = Tr(\rho^2)+1$, which was obtained by
Larsen \cite{Larsen90} and Ivanovic \cite{Ivanovic92};
here, $p_{i_m}=\bra{i_m} \rho \ket{i_m}$
denotes the probability of obtaining the $i$th result when projecting the state onto
the $m$th MUB.

We shall first extend this equality to an inequality valid in the case of a number $M$ of MUBs on a Hilbert space of arbitrary dimension.

\begin{theorem} \label{lem11}
Suppose $\rho$ is the state of a $d$-dimensional qudit, and let
$p_{i_m}=\bra{i_m} \rho \ket{i_m}$
denote the probability of obtaining the $i$th result when projecting the state onto
the $m$th MUB. If $M$ such MUBs exist, we have
\bgeq
\sum_{m=1}^M \sum_{i=1}^d p_{i_m}^2 \leq Tr (\rho^2) + \frac{M-1}{d} .  \label{probuncer1}
\edeq
\end{theorem}

{\bf Proof.}
For the sake of the proof, consider two qudits $a$ and $b$ and a basis of the composite system $ab$ that contains the following
$M(d-1)+1$ orthonormal basis states
\bgeq
|\Phi\rangle_{ab}=\frac 1{\sqrt d}\sum_{i=1}^{d}|i_1\rangle_a\otimes |i_1\rangle_b^*, \quad
|\phi_{m,k}\rangle_{ab}=\frac 1{\sqrt d}\sum_{i=1}^{d}\omega^{k(i-1)}|i_m\rangle_a\otimes |i_m\rangle_b^*
\label{basis11}
\edeq
with $\omega=e^{2\pi i/d}$, $k=1,\ldots, d-1$, and $m=1,2,\ldots,M$ and with the remaining basis states denoted as
$|\alpha_l\rangle_{ab}$ $(l=1,2,\ldots,
L=d^2-M(d-1)-1)$. Here $|i_m\rangle^*$ denotes the ``time-reversed" state of
$|i_m\rangle$; i.e., for a definite basis, say the first one $\{|i_1\rangle\}$,
the basis vectors coincide, $|i_1\rangle=|i_1\rangle^*$,
while all other bases differ by a complex conjugation of their expansion coefficients
on the first basis.

Given any density matrix $\rho$ of our single qudit, we define a two-qudit pure state $\rho_a\otimes I_b|\Phi\rangle_{ab}$
and expand it under the basis defined above
\bgeq
\rho_a\otimes I_b|\Phi\rangle_{ab}=\frac1d|\Phi\rangle_{ab}+\sum_{m=1}^M\sum_{k=1}^{d-1}\rho_{mk}|\phi_{m,k}\rangle_{ab}+\sum_{l=1}^{L}c_l|\alpha_l\rangle_{ab}
\label{expansion11}
\edeq
where $\rho_{mk}=\frac1d\sum_{i=1}^{d}\omega^{-k(i-1)}p_{i_m}$ with $p_{i_m}=\langle i_m|\rho|i_m\rangle$. A straightforward calculation yields
\bgeq
\sum_{m=1}^M\sum_{k=1}^{d-1}|\rho_{mk}|^2=\frac1{d^2}\sum_{m=1}^M\sum_{k=1}^{d-1}\sum_{i,j=1}^{d}\omega^{-k(j-i)}p_{i_m}p_{j_m}
=\frac1{d^2}\sum_{m=1}^M\sum_{i,j=1}^{d}(d\delta_{ij}-1)p_{i_m}p_{j_m}=\frac1d\sum_{m=1}^M\sum_{i=1}^{d}p_{i_m}^2-\frac M{d^2}.
\edeq
Thus
\bgeq
_{ab}\langle\Phi|\rho^2\otimes I|\Phi\rangle_{ab}=\frac1d Tr (\rho^2)\ge\frac 1{d^2}+\sum_{m=1}^M\sum_{k=1}^{d-1}|\rho_{mk}|^2
=\frac1d\sum_{m=1}^M\sum_{i=1}^{d}p_{i_m}^2-\frac {M-1}{d^2}
\edeq
i.e.,
\bgeq
\sum_{m=1}^M\sum_{i=1}^{d}p_{i_m}^2 \leq Tr(\rho^2)+\frac {M-1}{d}.
\edeq
This completes the proof.

Note that our construction of the two-qudit states resembles the
Jamiolkowski isomorphism \cite{Jamiolkowski}, and the expansion of
$(\rho \otimes I) |\Phi\rangle_{ab}$ in (\ref{expansion11}) can
indeed be viewed as an expansion of the operator $\rho$ in terms of
a set of orthonormal unitary operators in the Hilbert-Schmidt space
of operators, including $\{ I, U_{m,k}=\sum_{i=1}^d \omega^{k(i-1)}
\ket{i_m}\bra{i_m} | k=1,\cdots,d-1, \textrm{ and } m=1,\cdots,M \}$.

\begin{theorem} \label{popenuncert}
Following the same notation as above, we have the following entropic uncertainty relation for $M$
MUBs of a qudit system in the state $\rho$,
\bgeq
\sum_{m=1}^{M} H\{p_{i_m};i\} \geq
a C (K+1) \log_2 (K+1) + (1-a) C K \log_2 K
\label{the401}
\edeq
where $K= \lfloor \frac{M}{C} \rfloor$, $a=\frac{M}{C}-K$,
and $C$ has to be an upper bound for $\sum_{m=1}^M\sum_{i=1}^{d}p_{i_m}^2$.
We can, for example, use (\ref{probuncer1}) to choose $C= Tr(\rho^2)+\frac{M-1}{d}$.
\end{theorem}

{\bf Proof.}
Our proof uses a result
by Harremo\"{e}s and Tops\o e \cite{HT01}, which is conveniently
formulated as following.

{\sl Harremo\"{e}s-Tops\o e theorem.} For any given probability distribution
$\mathbf p=(p_1,p_2,\ldots,p_d)$, the Shannon entropy $H\{p_i;i\}$ and
the so-called index of coincidence, $C\{p_i;i\}=\sum_ip_i^2$, obey the
following inequality for arbitrary values of the
integer $1\le k\le d-1$:
\bgeq
H\{p_i;i\} \ge
\left((k+1)\log_2 (k+1) - k \log_2 k \right)
- k (k+1) \left(\log_2 (k+1) - \log_2 k \right) C\{p_i; i\} .
\edeq
As a result
\bgeq
\sum_{m=1}^M H\{p_{i_m};i\}\ge
M \left((k+1)\log_2 (k+1) - k \log_2 k \right)
- k (k+1) \left(\log_2 (k+1) - \log_2 k \right) \sum_{m=1}^M C\{p_{i_m};i\} .
\edeq
Since $\sum_{m=1}^M C\{p_{i_m};i\} \leq C $, the upper bound for $\sum_{m=1}^M\sum_{i=1}^{d}p_{i_m}^2$,
we immediately get
\bgeqn
\sum_{m=1}^{M} H\{p_{i_m};i\} &\geq&  M \left( (k+1) \log_2 (k+1) -k \log_2 k \right)
- k (k+1) \left( \log_2 (k+1) - \log_2 k \right) C  \nonumber \\
&=& (M-kC) (k+1) \log_2 (k+1) -(M-(k+1)C) k \log_2 k
\label{the41}
\edeqn
for any integer $k$ with $1\leq k \leq d-1$.

The right hand side of the above inequality can be viewed as a function of
the integer $k$, which reaches its maximal value at
$k=\lfloor M/C \rfloor$ when $M/C$ is not an integer and
which reaches the maximal value at both $k=\lfloor M/C \rfloor$ and
$k=\lfloor M/C \rfloor -1$ when $M/C$ is an integer
(see Appendix A).
Therefore, if we let $k=K=\lfloor M/C \rfloor$, we immediately get (\ref{the401}), which is the strongest inequality we can get from (\ref{the41}).
This completes the proof of Theorem \ref{popenuncert}.

\section{New entropic uncertainty relations}

The uncertainty relations, cited in Sec. 2 were all valid independently of the
state occupied by the physical system. Using our propositions, we can derive
state-dependent uncertainty relations, which must be obeyed for any MUBs for
a given state $\rho$, and we can use our results to derive also general state
independent uncertainty relations.

\begin{proposition}
For $M$ MUBs of a qudit prepared in the state $\rho$, we have the following
simple state-dependent entropic uncertainty inequality:
\bgeq
\sum_{m=1}^{M} H\{p_{i_m};i\} \geq  M \log_2   \frac{M}{C}   \label{the413}
\edeq
with $C=Tr (\rho^2)+ \frac{M-1}{d}$. Using that $Tr(\rho^2) \leq 1$, we obtain from (\ref{the413}) the following state-independent entropic uncertainty
inequality:
\bgeq
\sum_{m=1}^{M} H\{p_{i_m};i\} \geq  M \log_2   \frac{Md}{d+M-1} .   \label{the415}
\edeq
\end{proposition}
{\bf Proof.} By denoting the right-hand side of (\ref{the401}) as $f(K)$,
from the convexity of the function $x\log_2 x$ we immediately have
\bgeq
f(K) \geq  C\left( a(K+1) +(1-a)K \right) \log_2 \left( a(K+1) +(1-a)K \right)
= M \log_2 \frac{M}{C}
\edeq
which implies (\ref{the413}). Furthermore, (\ref{the415}) follows from (\ref{the413}) since $Tr(\rho^2)\leq 1$.
(\ref{the413}) also follows directly from (\ref{probuncer1}) by the convexity of the function $-\log_2 x$.

When the number of MUBs is large compared with $\sqrt{d}+1$, or more precisely, when
$M>\left( Tr (\rho^2)- \frac{1}{d}\right)\frac{d}{\sqrt{d}-1}$, our relation (\ref{the413}) is stronger than (\ref{uncer2}).
Also, when $M$ is small compared with $\sqrt{d}+1$, (\ref{the413}) provides a stronger relation than (\ref{uncer2})
when the state $\rho$ is sufficiently mixed.

Going back to the inequality (\ref{the401}), and making use of $Tr(\rho^2) \leq 1$ to
choose $C=1+\frac{M-1}{d}$ here, we get a state independent inequality which is in fact
stronger than (\ref{the415}).

\begin{proposition}
\bgeq
\sum_{m=1}^{M} H\{p_{i_m};i\} \geq \big( a  (K+1) \log_2 (K+1) + (1-a)  K \log_2 K \big) \frac{d+M-1}{d}
\label{the414}
\edeq
with $K= \lfloor \frac{Md}{d+M-1} \rfloor$ and
$a=\frac{Md}{d+M-1} -K$. The inequality (\ref{the414}) can also be rewritten as
\bgeq
\sum_{m=1}^{M} H\{p_{i_m};i\} \geq
M \log_2 K
+ (K+1) \left( M- K \frac{d+M-1}{d} \right)
\log_2 \left( 1+ \frac{1}{K} \right)
\label{the41005}
\edeq
which is dominated by the first term when $M$ is much larger than unity.
\end{proposition}

As any system with $d\geq 2$ has at least $3$ MUBs, we will consider that case as an example, and note from (\ref{the414}) that
\bgeq
\sum_{m=1}^{3} H\{p_{i_m};i\} \geq \left\{
\begin{array}{ll}
2 & \textrm{ for }d=2\textrm{; } \\
\frac{8}{3} & \textrm{ for }d=3\textrm{;} \\
3 (1-\frac{4}{d}) \log_2 3 +\frac{12}{d}  & \textrm{ for }d\geq 4\textrm{.}
\end{array}
\right.
\edeq

Unlike the restrictions on previously derived inequalities, the entropic uncertainty inequalities, derived here, work for any dimension $d$ of the system and any number $M$ of MUBs
(assuming they exist). When $d$ is a power of a prime, we know that there exist $d+1$ MUBs, and choosing $M=d+1$ in (\ref{the414}) we obtain the result in \cite{sanchez-ruiz95}
\bgeq
\sum_{m=1}^{d+1} H\{p_{i_m};i\}
\geq \left\{
\begin{array}{ll}
(d+1) \log_2 (\frac{d+1}{2}) &  \textrm{when $d$ is odd,} \\
\frac{d}{2} \log_2 (\frac{d}{2}) + (\frac{d}{2}+1) \log_2 (\frac{d}{2}+1) & \textrm{when $d$ is even.}
\end{array}
\right.
\edeq
Unlike the proof in \cite{sanchez-ruiz95}, which works only when $d$ is a power of a prime and $M=d+1$,
our result (\ref{the414}) works for any $d$ and any allowed number of MUBs $M$.

The state-dependent inequality with $C=Tr (\rho^2)+ \frac{M-1}{d}$ in (\ref{the401}) provides stronger bounds than (\ref{the413}) and (\ref{the414}).
Consider, for example, the qubit case $d=2$, and suppose $M=3$, with $C=Tr (\rho^2)+ 1$;
from (\ref{the401}), we have
\bgeq
\sum_{m=1}^{3} H\{p_{i_m};i\} \geq 4- 2 Tr (\rho^2). \label{qubituncert}
\edeq
This entropic uncertainty relation (\ref{qubituncert}) is stronger than the result
$\sum_{m=1}^{3} H\{p_{i_m};i\} \geq 2$ in \cite{sanchez-ruiz95,sanchez93}, and
it is also stronger than $\sum_{m=1}^{3} H\{p_{i_m};i\} \geq 3 \log_2 \frac{3}{1+Tr(\rho^2)}$ that
follows from (\ref{the413}).

\bigskip

{\bf Remark.} Inequality (\ref{probuncer1}) itself can be viewed as an entropic uncertainty relation in terms of
the Tsallis entropy, which is defined as
$S_q^T \{p_i; i\} \equiv \left( 1- \sum_i p_i^q \right)/(q-1)$ ($q > 1$) \cite{Tsallis},
with $q=2$ for
our case.  Similarly we can obtain inequalities obeyed by the $q=2$ R\'enyi entropy, defined by
$S_q^R \{p_i; i\} \equiv \log_2 \left( \sum_i p_i^q \right)/(1-q)$ ($q > 1$) \cite{Renyi}.
Using the concavity property of the R\'enyi entropy and setting $q=2$,
we get from (\ref{probuncer1}) the inequality
\bgeq
\sum_{m=1}^{M} S_2^R \{p_{i_m}; i\} \geq - M \log_2 \left( \frac{1}{M} (Tr \rho ^2 +\frac{M-1}{d}) \right)
\geq M \log_2 \frac{Md}{d+M-1}.
\edeq

Let us finally consider the application of entropic uncertainty to composite systems.
Let $A$  and $B$ denote subsystems with Hilbert space dimensions $d_A$ and $d_B$, and let $\{\ket{i_{mA}} | i=1,\cdots,d_A\}$ and $\{\ket{s_{nB}} | s=1,\cdots,d_B\}$
denote the $m$th and $n$th mutually unbiased bases of systems $A$  and $B$. We now
consider local measurements on a bipartite state $\rho_{AB}$ of the joint system.  When system $A$ is projected onto
the $m$th MUB and system $B$ is projected onto the $n$th MUB, the joint probability of outcomes in these bases is denoted by
$p^{(m,n)}_{is}=\bra{i_{mA}}\bra{s_{nB}} \rho_{AB} \ket{i_{mA}} \ket{s_{nB}}$.
The entropic uncertainty inequalities we have derived above can now be applied to the
composite system, and in particular we can derive the following.

If $\rho_{AB}$ is a separable state, then for $M$ MUBs of each subsystem we have
\bgeqn
\sum_{m=1}^M H\{ p^{(m,m)}_{is} ; is \} &\geq&
M \log_2 K_A + (K_A+1) \left( M- K_A \frac{d_A+M-1}{d_A} \right) \log_2 \left( 1+ \frac{1}{K_A} \right) \nonumber \\
&+&
M \log_2 K_B + (K_B+1) \left( M- K_B \frac{d_B+M-1}{d_B} \right) \log_2 \left( 1+ \frac{1}{K_B} \right)
\edeqn
with $K_{A(B)}= \lfloor \frac{Md_{A(B)}}{d_{A(B)}+M-1} \rfloor$.

{\bf Proof.} If $\rho_{AB}$ is separable, it can be written as a convex sum of product states:
$\rho_{AB} =\sum_j q_j \rho_j^A \otimes \rho_j^B$. Therefore we have
\bgeqn
\sum_{m=1}^M H\{ p^{(m,m)}_{is} ; is \}_{\rho}
&\geq& \sum_j  q_j  \sum_{m=1}^M H\{ p^{(m,m)}_{is} ; is \}_{\rho_j^A\otimes \rho_j^B} \nonumber \\
&=& \sum_j q_j \sum_{m=1}^M H\{ p^{(m)}_{i} ; i \}_{\rho_j^A} +\sum_j q_j \sum_{m=1}^M H\{ p^{(m)}_{s} ; s \}_{\rho_j^B} .
\edeqn
The proposition immediately follows from the above inequality and (\ref{the41005}).

As an example, when $d_A=d_B=2$ and $M=3$, for a separable state $\rho_{AB}$ we have
\bgeq
\sum_{m=1}^{3} H\{ p^{(m,m)}_{is} ; is \} \geq 4 .
\edeq

It should be noted that this separability criterion is not a strong one, and  replacing the inequality sign by an equality it does not even for qubits provide the actual boundary between separable and entangled states. The result, however, is an example of how the concavity of entropy functions together with entropic uncertainty relations can provide insights into the topic of entanglement.

\section{Conclusion}

In this paper we have presented a number of inequalities obeyed by the probability
distributions for measurements on quantum systems in mutually unbiased bases.
We have obtained tighter and more general entropic uncertainty relations than
the ones presented in the literature, and we have given less tight, but more compact
expressions in simple cases. In the Introduction we motivated the work by the application
of complementarity and uncertainty relations in quantum information theory.
Entropy is used to quantify information, and hence entropic uncertainty relations
provide bounds on the information obtainable by measurements of different observables
of a quantum system. The more general inequalities derived and proven in this article
thus form the basis for new quantitative results on this topic.

\section*{Acknowledgments}
The authors wish to thank Uffe V. Poulsen and the MOBISEQ network under the Danish Natural Science Research Council
for helpful discussions.
S. W. and S. Y. also wish to acknowledge support from the NNSF of China (Grants No. 10604051 and 10675107),
the CAS, and the National Fundamental Research Program.

\section*{Appendix}

Denoting $f(x)=(M-C x)(x+1)\ln(x+1)-(M-C(x+1))x\ln x$ and $K=\left\lfloor\frac M C\right\rfloor$,
we shall prove $\max_k{f( k )}\le f( K)$. Here $k$ is any integer and $1\leq k \leq d-1$.
$g(x)=x\ln x - (1+x)\ln(1+x)$ is a decreasing function of $x\ge 0$ since $g^\prime(x)=\ln\frac x{1+x}<0$.
Thus $g(k+1)<g(k )$, i.e.
$$2\ln(k+1)^{k+1}-\ln k^k(k+2)^{k+2}<0$$
Denote $\Delta(x)=f(x)-f(x+1)$, which reads
$$\Delta(k )=(M-C(k+1))\left(2\ln(k+1)^{k+1}-\ln k^k(k+2)^{k+2}\right)$$

If $\left\lfloor\frac M C\right\rfloor$ is not an integer,
then $\frac M C -1 < K < \frac M C$
and $\Delta(K-1) <0$, so $f(K-1) < f(K)$, and $\Delta(K) >0$, so $f(K) > f(K+1)$. So the maximal
value of $f(k)$ over integer $k$ is obtained at $k=K$.
If $\left\lfloor\frac M C\right\rfloor$ is an integer, then $ K = \frac M C$ and
$\Delta(K-1) =0$, so $f(K-1) = f(K)$; similarly, we can show $f(K-2) < f(K-1)$ and $f(K+1) < f(K)$.
So the maximal value of $f(k)$ over integer $k$ is obtained at both $k=K$ and $k=K-1$.
Therefore $\max_k{f( k )}\le f( K)$.

\end{document}